\documentclass[aps,prd,superscriptaddress,twocolumn,floatfix,nofootinbib]{revtex4-1}
\usepackage{graphicx}  
\usepackage{dcolumn}   
\usepackage{bm}        
\usepackage{amssymb}   
\usepackage{amsmath}
\usepackage{epsfig}
\usepackage{rotating}
\usepackage{color}
\newcommand{\be}{\begin{equation}}
\newcommand{\ee}{\end{equation}}

\newcommand{\ba}{\begin{eqnarray}}
\newcommand{\ea}{\end{eqnarray}}
\renewcommand\[{\begin{equation}}
\renewcommand\]{\end{equation}}
\begin{document}



\title{ Do massive compact objects without event horizon exist in infinite derivative gravity? }

\author{Alexey S. Koshelev}
\affiliation{Departamento  de F\'isica and Centro  de  Matem\'atica  e 
Aplica\c c\~oes,  Universidade  da  Beira  Interior,  6200  Covilh\~a, 
Portugal}
\affiliation{Theoretische Natuurkunde, Vrije Universiteit Brussel and The 
International Solvay Institutes, Pleinlaan 2, B-1050, Brussels, Belgium}

\author{Anupam Mazumdar}
\affiliation{Van Swinderen Institute, University of Groningen, 9747 AG, Groningen, The Netherlands}

\begin{abstract}
Einstein's General theory of relativity is plagued by cosmological and blackhole type singularities Recently, it has been shown that
infinite derivative, ghost free, gravity can yield non-singular cosmological and mini-blackhole solutions. In particular, the theory 
possesses a {\it mass-gap} determined by the scale of new physics. This paper provides a {\it plausible  argument}, not a no-go theorem, based on 
 the {\it Area-law} of gravitational entropy that within infinite derivative, ghost free, gravity non singular compact objects in the static limit need not have  horizons.
\end{abstract}

\maketitle

\section{Introduction}
Einstein's theory of general relativity (GR) is extremely successful theory in the infrared (IR)~\cite{Will:2014kxa}, matching all the 
current observations including predictions of gravitational waves from coalescing binary blackholes~\cite{Abbott:2016blz}. The only 
drawback arises in its predictions in the ultraviolet (UV) regime, where there exists classical singularities, such as blackhole and 
cosmological Big Bang singularities, see~\cite{Hawking:1973uf,Wald:1984rg}. There exists a vicious circle in GR, which
inevitably leads to a collapse of a normal matter satisfying all the known energy conditions~\cite{Penrose:1964wq}. 

Since in GR both energy and pressure gravitates, therefore, any normal matter satisfying  strong, weak and null energy conditions will always
lead to a focusing of either time-like and/or null rays according to the Raychoudhury's equation~\cite{Raychaudhuri:1953yv}, which will lead to a formation 
of a trapped surface, and an apparent horizon, see~\cite{Frolov-book}. In the static limit, both apparent and event horizon coincides, and the metric potential is given by the Schwarzschild metric, which asymptotes to the Minkowski far away from the source. The potential is denoted by $\Phi \sim G m/r$, where $m$ is the mass of a blackhole, and $G$ is the Newton's constant~\footnote{We will also use the $4$ dimensional gravitational Planck constant: $M_p=1/\sqrt{G}\sim 10^{19}$~GeV.}. In the Schwarzschild's geometry, where  $ r\leq 2Gm$, the metric potential keeps growing all the way to $\Phi \rightarrow \infty$ as $r\rightarrow 0$. 

In the cosmological context, in the homogeneous and isotropic Universe,  although the origin of singularity is slightly different than 
that of a blackhole, but there is one common ingredient, that is the energy density blows up as $t\rightarrow 0$, and so is the metric potential.

The aim of this paper is to break this vicious circle of inevitable collapse of matter in gravity without violating any of the energy conditions. 
Our arguments will be based on a static case in this paper, so strictly speaking they will not be directly applicable to a cosmological setup.
There are two ways we can try to avoid inevitable singularity in a gravitational theory:
(a) First option is to make gravity repulsive at short distances and small time scales, in the UV, such that a delicate 
balance of matter and gravitational pressure would halt the collapse problem, and therefore avoiding singularity.  This approach will inevitably lead 
to introduction of ghosts in the gravitational sector.
(b) The second option is to make gravitational interaction sufficiently weak, such that in the UV the spacetime becomes regular and the gravitational 
force between particles vanishes, $F_g \rightarrow 0$, therefore  the gravitational binding energy ceases to be singular within a finite length and time scale.
We will follow the second strategy here.  

Here we will not attempt to provide a mathematical proof for avoiding either singularity or event horizon for systems as massive as
astrophysical solar massive objects, but will present some arguments of plausibility regarding the possible static properties of non-singular compact objects.

\section{Ghost free infinite derivative gravity}

We wish to seek a theory of gravity, which allows weak field limit in the entire region of spacetime, i.e. $ 2\Phi  < 1$ for 
all values of $r$. In the static and spherically symmetric geometry, this would mean $\Phi < 1$ for both $r\geq 2GM$ and $r\leq 2GM$.
Indeed, at sufficiently large values, we would like to recover $1/r$-fall of Newtonian potential. Therefore, the desirable space-time metric should
be  the linearized form always, specifically in a static geometry:
\begin{equation}\label{met}
ds^2 = - (1-2\Phi) dt^2+ (1+2\Phi)dr^2\,,
\end{equation}
where $dr^2=dx^2+dy^2+dz^2$, and $2\Phi < 1$, such that the solution remains perturbative for a given modification of GR.

It has been shown recently that {\it infinite derivative gravity} (IDG) can avoid cosmological and blackhole singularities by making the gravitational action
{\it ghost free}~\cite{Biswas:2011ar,Biswas:2005qr}, such that the propagating degrees of freedom remain massless spin-2 and spin-0 
components~\cite{Tomboulis, Biswas:2011ar,Biswas:2005qr, Modesto,Biswas:2013kla}.  In addition to this, 
IDG also allows massless spin-2 and  one massive spin-0 components to propagate, analogous to the Brans-Dicke gravity, see~\cite{Biswas:2013kla,Biswas:2016egy}. The most general 
ghost free IDG action  can be recast as~\cite{Biswas:2011ar,Biswas:2016egy}:
\begin{eqnarray}\label{action}
S= \int d^4x \sqrt{-g}\left [ M_p^2R + R{\cal F}_1(\bar\Box)R+R^{\mu\nu}{\cal F}_2(\bar\Box)R_{\mu\nu}\right. \nonumber \\
\left. +R^{\mu\nu\lambda\sigma} {\cal F}_{3}(\bar \Box)R_{\mu\nu\lambda\sigma}\right]\,,
\end{eqnarray}
where $\bar\Box =\Box/M_s^2$, and $M_s$ is the new scale of physics which appears below the $4$ dimensional Planck mass, i.e. $M_s\leq M_p$.
It has been shown in Refs.~\cite{Tomboulis,Tseytlin:1995uq,Siegel,Biswas:2014yia,Biswas:2004qu,Talaganis:2014ida} that $M_s$ harbours the scale 
of non-locality at the level of quantum interactions.

The three gravitational form-factors behave as $2{\cal F}_1+{\cal F}_2+2{\cal F}_3=0$ around a Minkowski background~\cite{Biswas:2011ar}, in order to propagate only the massless graviton.  In order to avoid any new dynamical degrees of freedom, that includes tachyons and/or ghosts.
the propagator must be suppressed by {\it exponential of an entire function}~\cite{Biswas:2011ar,Tomboulis}. 
\begin{equation}\label{propagator}
\Pi(k^2)\sim e^{-\gamma(k^2)}\left[\frac{P^{(2)}}{k^2}-\frac{P^{(0)}}{2k^2}\right]\,,
\end{equation}
where $P^{(2)},~P^{(0)}$ are spin-projection operators respectively. 
The exponential factor with $\gamma(k^2)$ being an entire function is mathematically an unique option  in order to avoid new poles at finite momenta. Surely, $\gamma(k^2)$ must obey certain conditions. In particular, it should grow at large momenta such that $\gamma(k^2)$ decays in the UV, signalling the weakening of the graviton propagation for physical degrees of freedom in UV, and in the limit when $k\rightarrow 0$, in the IR the propagator matches exactly the behaviour of 
Einstein's gravity. The simplest examples of such an entire functions are polynomials, and we concentrate our discussions with $\gamma(k^2)=k^2/M_s^2$. Note that in \cite{Edholm:2016hbt} other polynomials and generic series for function $\gamma(k^2)$ were analyzed and proven to be compatible with the theory. However, the main idea of the current paper can be presented by using just the simplest monomial, and moreover, conceptually our conclusions will not be affected by considering a more general form of $\gamma(k^2)$.

In the limit when $M_s\rightarrow \infty$, the theory 
comes back to the predictions of the IR, i.e. the Einstein-Hilbert action~\footnote{Such exponential suppression in the propagator 
is quite common in theories with infinite derivatives, whether they arise
in a bosonic sector,  or in a fermionic sector~\cite{Evens:1990wf,Biswas:2014yia,Evens:1990wf}. In fact, such a propagator also arises in string field theory~\cite{Witten:1985cc,Freund:1987kt,Frampton:1988kr,Siegel:1988yz}.  }.
The above modified propagator ensures that in the UV, the propagator is exponentially suppressed, thus improving upon the UV 
behaviour of gravity.  The vertex operator on the other hand gets exponential enhancement, therefore at a quantum level, the interactions 
become non-local. The superficial degree of divergence suggests that such gravitational theory becomes power counting renormalizable 
for loop $L>1$\cite{Tomboulis,Modesto,Talaganis:2014ida}.

\section{Gravitational potential from a source}

A very interesting consequence of this suppressed propagator can be illustrated by the fact that for a point-source, the 
linearized equations of motion yields~\cite{Biswas:2011ar}:
\begin{equation}\label{eqm}
e^{-\Box/M_s^2}\Box \Phi=8\pi G\rho= m\delta^3(r)\,.
\end{equation}
In the static limit, when $\Box =\nabla^2$
 the gravitational potential can avoid $1/r$ singularity, which is impossible to avoid otherwise in GR. The metric potential has a solution given by~\cite{Biswas:2011ar}:
\begin{equation}\label{soln}
\Phi(r) = \sqrt{\frac{\pi}{2}}\frac{m}{M_p^2 r} {\rm erf}(M_s r/2) \,,
\end{equation}
where for $r\gg 2/M_s$, the potential falls as the Newtonian limit: $1/r$, and for $r< 2/M_s$, the potential 
asymptotes to a constant value, and the gravitational force vanishes:
\begin{equation}\label{const}
\Phi \sim mM_s/M_p^2< 1\,,~~F_g\rightarrow 0\,,~~{\rm for}~r<\frac{2}{M_s}\,,
\end{equation}
and for mass enclosed within such a radius follows: $m < M_p^2/M_s$. 
For $r> 2/M_s$, the ${\rm erf}\rightarrow 1$, and therefore one recovers the standard $1/r$-fall of gravity.
As it has been shown earlier in \cite{Frolov:2015bia,Edholm:2016hbt},  a constant asymptotic in the region $r < 2/M_s$ and standard $1/r$ fall-off at large distances are the universal features of IDG, while the value of the  cross-over scale $2/M_s$ depends on the particular form of function $\gamma(k^2)$ in Eq.~(\ref{propagator}). Namely, the factor of $2$ in the cross-over scale is subject to the lowest degree of $k^2$ in the series expansion of $\gamma(k^2)$. For, $\gamma(k^2)\approx(k^2/M_s^2)^n+O(k^{2n+2})$ with $n>1$ the factor of $2$ will be subject to modification. In spite of these details, let us stick with the lowest order polynomial for the purpose of our discussion.

In this regard the 
IDG possess a scale, known as {\it mass gap}, determined by the scale of non-locality $M_s$~\cite{Biswas:2011ar,Frolov:2015bta}~\footnote{
This has now been verified by Valeri Frolov  and his collaborators in a dynamical context as well, see Refs.~\cite{Frolov:2015bia,Frolov:2015bta,Frolov:2016pav}.
Note however that there are results~\cite{Li:2015bqa}, which contradict the results even at the linearized limit~\cite{Biswas:2011ar,Frolov:2015bta,Frolov:2015bia,Frolov:2015bta,Frolov:2016pav}. These papers~\cite{Li:2015bqa} do not attempt to solve the complete equations of motion for the ghost free IDG, the complete equations of motion have been derived in Ref.~\cite{Biswas:2013cha}. There is another  difference, in our case we have an explicit source term, Eq.~(\ref{eqm}), which is apparently lacking in their solution. It has been assumed that a vacuum solution will be similar to the Schwarzschild, but there is no explicit proof given.}.

The current constraint on $M_s$ arises from table-top
laboratory experiment, which has seen no deviation from Newtonian gravity up to $5.6\times 10^{-5}$~m~\cite{Kapner}. This limit 
translates to: $M_s\geq 0.004$~eV~\cite{Edholm:2016hbt}, and in order for $\Phi < 1$, we obtain a bound on mass
\begin{equation}
m\leq \frac{M_p^2}{M_s}\sim 10^{25}~{\rm grams}\,,
\end{equation}
which is roughly the mass of the Moon. Therefore, for a Moon like massive system would never generate a metric singularity 
at $r=0$, and its potential will be regular all the way without forming any event horizon, i.e. $2\Phi < 1$. If we had 
chosen $M_s\sim M_p$, the bound on $m$ would be:
\begin{equation}
m\leq M_p\sim 10^{-5}~{\rm grams}\,,
\end{equation}
ideal for a planckian size object, i.e. $r \sim 2 M_p^{-1} $, where the metric potential is constant. Such regions of space-time, with a 
constant metric potential, could be treated as ``{\it plaquette}",  which we denote here by ${\cal U}$.

The property of a {\it plaquette} is solely determined by $M_s$ and $M_p$ within IDG.  If we had chosen $M_s=10^{16}$~GeV, then such  a {\it plaquette}
would be able to hold  $m\leq M_p^2/M_s \sim 10^{-2}$~grams, without forming a singularity within $r\leq 2/M_s$. Strictly speaking,  this notion holds true for a static geometry and some of these arguments will modify in a time dependent case~\footnote{Many authors have considered latticising the space-time in the context of quantum gravity, see \cite{Bombelli:1987aa,Rovelli:1994ge}, for a review see~\cite{Ambjorn:2012jv}.}.

\section{Non-singular compact object : space-time filling {\it plaquettes}}

The key point to note here is that the  {\it plaquette's} mass is determined by the mass gap $M_s$. Since 
the planckian energy density is the largest one can achieve, there exists a simple bound on a
{\it plaquette} with a constant potential from Eq.~(\ref{const}):
\begin{eqnarray}\label{ed0}
 m^2M_s^2 \leq M_p^4\,,~~{\rm for}~~r<\frac{2}{M_s}\,.
\end{eqnarray}
In the static limit, even if there are $N$ such {\it plaquettes}, within a compact region of space-time,
one would naturally expect the largest energy density to be  still given by the planckain one:
\begin{eqnarray}\label{cond}
N^2m^2\left(\frac{M_s}{\sqrt{N}}\right)^2&\equiv & M_{ns}^2M_{eff}^2\leq M_p^4\,,\nonumber \\
{\rm for}~~~&& r <r_{\ast}=\frac{2}{M_{eff}}\,,
\end{eqnarray}
where  the rest mass of a non-singular compact object (NSCO) is now given by: $M_{ns} = Nm$. Note that in order for the above condition to hold true, 
the effective non-local scale has shifted from $M_s$ to a much lower value:
\begin{equation}
M_{eff}\sim \frac{M_s}{\sqrt{N}}\,.
\end{equation}
A simple but compelling  argument can be presented from the {\it entropy} point of view. Any gravitationally
bound system is known to possess a {\it gravitational entropy}~\cite{Bekenstein:1973ur,Hooft,Wald-1,Abreu:2010sc}.

Let us imagine that the gravitationally bound space-time, ${\cal M}$, is filled up with such constant potential {\it plaquettes}, ${\cal U}$, such 
that the gravitational potential is constant over ${\cal M}$ on a macroscopic scale, i.e. $r_\ast = 2/M_{eff}$.
If every {\it plaquette}, ${\cal U}$, of size $\sim 2/M_s$ denotes 1-unit of entropy, then the space-time {\it filling} $N$ such constant potential {\it plaquettes}
will have entropy scaled by $N$-units. One would expect similar result from computing the gravitational entropy for 
a gravitationally bound system, ${\cal M}$, with a constant potential $\Phi< 1$ inside $r\leq r_\ast$.

Note that the leading order contribution to the gravitational entropy is given by the {\it Area-law},  
where $S_g={\it Area}/4G$, where ${\it Area}$ is the enclosed area of such a bound system, and $G$ is the Newton's 
constant~\cite{Bekenstein:1973ur,Hooft,Wald-1,Abreu:2010sc}. At the leading order the {\it Area-law} holds true for the
ghost free IDG as well, see~\cite{Conroy:2015wfa}. As long as the metric potential, $\Phi < 1$, the Wald's gravitational entropy 
comes out to be proportional to the ${\it Area}$ for the static and spherically symmetric bound system.

Therefore, in our case, the gravitational entropy for ${\cal M}$ would scale as $\propto 4\pi r_{\ast}^2$, where
$r_\ast = \sqrt{N}/M_s$:
\begin{equation}\label{entropy}
S_g\equiv \frac{{\rm Area}}{4G} \propto N\left(\frac{M_p}{M_s}\right)^2\sim \left(\frac{M_p}{M_{eff}}\right)^2\,.
\end{equation}
This analysis suggests that for a system with a mass-gap, there is a possibility to shift the scale of non-locality from 
$M_s$ to $M_{eff}=\sqrt{N}/M_s$ for a gravitationally bound system, i.e. $\Phi \sim{\rm constant}$ within radius 
$r_{\ast}= 2/\sqrt{M_{eff}}$~~\footnote{This scaling of non-locality can also be seen as a property of field theory with 
infinite derivatives at a quantum level. It has been shown that the scalar counterpart of Eq.~(\ref{action}), which preserves the scaling and the 
shift symmetry, see~\cite{Talaganis:2014ida},  has a similar property in the scattering amplitudes. The scattering amplitude is exponentially suppressed
\cite{Talaganis:2016ovm,Biswas:2014yia} in the UV, when centre of mass energy exceeds $M_s$. However, when multiple scatterings are considered, the
scattering amplitude becomes suppressed by the new {\it effective} scale $M_{eff} \rightarrow M_s/\sqrt{N}$ for large $N$-limit~\cite{Talaganis:2017dqy}. }.

Based on this analysis we can now ask; could we form a super-massive NSCO
with a radius  bigger than the Schwarzschild's radius, $r_{sch}$, i.e. $r_\ast\geq r_{sch}$.
Within IDG it is indeed possible to construct  ${\cal M}$, which has  a constant metric potential.
In fact, following the arguments of the last section, we can now find the metric potential  to be:
\begin{equation}\label{new-soln}
\Phi(r) = \sqrt{\frac{\pi}{2}}\frac{M_{ns}}{M_p^2 r} {\rm erf}(r M_{eff} /2) \,,
\end{equation}
where $M_{ns}=Nm$ is the mass of NSCO. The solutions of the above equation yields:
\begin{equation}\label{metric-new}
\Phi \sim \left\{
      \begin{array}{ll}
          \frac{M_{ns}M_{eff}}{M_p^2} < 1,
          & r< r_{\ast}=\frac{2}{M_{eff}}, \\[3mm]
          \frac{GM_{ns}}{r} < 1, & r> r_{\ast}=\frac{2}{M_{eff}}.
      \end{array}
      \right.
\end{equation}
Now, we can seek under what conditions  $r_{\ast}\geq r_{sch}$: 
 \begin{equation}\label{cond-1}
r_{\ast}= \frac{2}{M_{eff} }\geq r_{sch} =\frac{2M_{ns}}{M_p^2}~~\Longrightarrow~~ M^2_{ns}M^2_{eff} \leq M_p^4\,,
 \end{equation} 
which is a similar condition as that of the scaling of the entropy argument, see Eq.~(\ref{entropy}), and also Eq.~(\ref{cond}).

We can also estimate what should be the value of $N$ if NSCO were made up of a billion times the solar mass, i.e. $10^{12}\times 10^{33}\sim 10^{45}$~grams.
For $M_s =M_p$, $N\geq 10^{100}$, while, for $M_s= 10^{16}$~GeV,  $N\geq 10^{94}$, and for $M_s = 10^{4}$~GeV, $N\geq 10^{70}$. Indeed, 
a  NSCO can hold a large $N$, which signifies large amount of entropy, very similar to the case of  a typical blackhole within GR.

 \section{Absence of an event horizon}
 
One of the properties of NSCO within IDG is that the absence  of an event horizon, since $2\Phi < 1$ in the macroscopic region of space-time, as large
as that of the Schwarzschild's radius, $r_\ast\geq r_{sch}$, see Eq.~(\ref{cond-1}). By construction
the metric potential is given by Eq.~(\ref{metric-new}), in conjunction with Eq.~(\ref{cond-1}).

Indeed Sagitarus A$^{\ast}$, which harbours a huge mass in the centre of our Milkyway~\cite{Doeleman:2008qh}, yields  $1/r$ gravitational potential for a distant observer, $r> r_{\ast}> r_{sch}$. In the idealised case, assuming in the static scenario, when the observer comes close to the vicinity of its Schwarzschild radius the gravitational potential tends to become constant for $r< r_{sch} < r_{\ast} $. The transition is indeed smooth and determined by the full solution, Eq.~(\ref{new-soln}).

The absence of an event horizon is indeed a very compelling reason why such NSCO should be further studied. The Hawking's information-loss paradox~\cite{Hawking:1974sw} can be resolved amicably in the absence of any stretched horizon. The absence of an event horizon means that the information can never get lost, but indeed, NSCO can modify how the information can be retrieved or scrambled due to large N {\it plaquettes}, which requires further study, see for instance ~\cite{Frolov:2017rjz}.
Here, we wish to draw some similarity with the fuzz-ball scenario~\cite{Mathur:2005zp}, where the physical situation has some common feature, though the intricate details are very different. In the fuzz-ball scenario there is no event  horizon either, and it also reiterates that the scale of quantum gravity need not be localised on a small scale, but it can be enlarged to a macroscopic length scale in a self gravitating system~\cite{Mathur:2005zp}. This happens purely due to stringy origin, i.e. presence of non-local objects such as winding modes, branes, etc. In our case, higher derivatives are akin to $\alpha'$ corrections in string theory, therefore there might be some interesting connection which one might be able to exploit between these two seemingly different approaches. Furthermore, besides fuzz-ball scenario, there are examples such as gravastars~\cite{Mazur:2004fk, Visser:2003ge}, and blackstars~\cite{Barcelo:2007yk,Carballo-Rubio:2017tlh,Barcelo:2014cla, Barcelo:2016hgb}, which mimic blackholes without event horizons.

\section{Conclusion}  
In the Einstein-Hilbert gravity, it is a challenge to avoid $1/r$ singularity in the metric potential for a static and spherically symmetric background.  
The IDG provides a compelling candidate of quantum gravity where the gravitational potential, $\Phi$, can be made {\it constant} in the UV, such that at short distances the gravitational force vanishes. The scale is typically determined by the mass gap, i.e. the scale of non-locality $M_s$. The property of such a region of space-time can be constructed purely by $M_s$ and $M_p$, here we have denoted them by {\it plaquette}. 

Although our {\it current} analysis does not  provide a {\it proof} of the absence of horizons in the ghost free IDG, but provides an argument that within IDG non-locality can be shifted to a larger length scales by introducing large number of {\it plaquettes}, or degrees of freedom, which is expected to be present in any astrophysical super-massive object.
 For $N$ such  {\it plaquettes} the scale of non-locality can be shifted to the IR from $M_s$ to  $M_s/\sqrt{N}$. For a large enough $N$, the gravitational potential, $\Phi$, inside the Schwarzschild volume can be made constant, where the gravitational force vanishes between particles. The entire region of space-time becomes weakly coupled bound state where there is no singularity in the metric potential. Outside this region, the metric potential behaves as $1/r$-fall in the static limit. Since, by construction the metric potential in the entire space-time remains bounded by unity, there is no formation of a trapped surface or an event horizon. The absence of an horizon in such a system has a benefit of addressing the information-loss paradox, which exists in the context of an event horizon.

In this context, the super heavy astrophysical objects can be thought of as NSCO. For instance, billion times solar massive objects can now be treated as NSCO
without an event horizon within IDG. Such an astrophsyical object will be slightly less compact than the standard blackhole within pure GR. In our case, the compactness parameter is determined by $r_\ast\geq r_{sch}$, which would be a smoking gun astrophysical signature for such objects. Furthermore, one can discuss shadow of an event horizon, quasi-normal ring-down phase of gravitational waves, etc. All these interesting questions, see~\cite{Sakai:2014pga,Cardoso:2016rao,Barcelo:2017lnx},
 would require further study in the context of IDG.
Finally, our approach here is valid for a static limit. It would be interesting to show how some of these arguments can be modified to study the dynamical setup,
such as in the case of a collapsing thick spherical shell in a relativistic, and a non-relativistic limit.

\section{Acknowledgement}
The authors would like to thank Tirthabir Biswas, Tomi Koivisto, Valeri Frolov, Gaetano Lambiase, Terry Tomboulis and Shinji Mukohyama for discussing various aspects of this problem. The authors would also like to thank the anonymous referee for a very helpful discussion, and for numerous suggestion to help improving the quality of the paper.
AK is supported by FCT Portugal investigator project IF/01607/2015, FCT
Portugal fellowship  SFRH/BPD/105212/2014 and in part by FCT Portugal
grant  UID/MAT/00212/2013.


\end{document}